\documentclass[5p,times,twocolumn]{elsarticle}

\usepackage{amssymb}

\usepackage{hyperref}
\makeatletter
    \g@addto@macro{\UrlBreaks}{\do\/\do\-\do\_}
\makeatother
\journal{}

\begin{document}

\begin{frontmatter}



\title{Quantifying national space heating flexibility potential at high spatial resolution with heating consumption data}

\author[Oxford]{Claire Halloran\corref{cor1}}

\cortext[cor1]{Corresponding author. Email address: claire.halloran@eng.ox.ac.uk}

\affiliation[Oxford]{organization={Department of Engineering Science, University of Oxford},
            addressline={Parks Road}, 
            city={Oxford},
            postcode={OX1 3PJ}, 
            country={UK}}
\author[Oxford]{Jesus Lizana}

\author[Oxford]{Malcolm McCulloch}

\begin{abstract}

Decarbonizing the building stock in cold countries by replacing fossil fuel boilers with heat pumps is expected to drastically increase electricity demand. While heating flexibility could reduce the impact of additional demand from heat pumps on the power system, characterizing the national spatial distribution of heating flexibility capacity to incorporate into sophisticated power system models is challenging. This paper introduces a novel method for quantifying at large scale and high spatial resolution the energy capacity and duration of heating flexibility in existing building stock based on historical heating consumption data and temperature data. This method can reflect the geographic diversity of the national building stock in sophisticated power system models. The proposed heating consumption-based method was tested in Britain using national residential gas data and was validated by comparing it with an Energy Performance Certificate-based method and an indoor temperature-based method. The results demonstrate the potential of this approach to characterize the heterogeneous distribution of heating flexibility capacity at the national scale. Assuming a 3ºC temperature flexibility window, a total thermal energy storage capacity of 500 GWh$_{th}$ is identified in the British residential housing stock. The electricity storage equivalent of this figure depends on the temperature-dependent coefficient of performance (COP) of the heat pumps: for an illustrative cold weather COP value of 2.5, this thermal energy storage capacity is equivalent to 200 GWh of electricity storage. Regarding heating flexibility duration, gas-heated homes have a median of 5.9 heat-free hours for 20th percentile regional daily winter temperatures from 2010 to 2022. However, extreme cold days nearly halve flexibility duration to a median of 3.6 heat-free hours. These high spatial resolution energy capacity and self-discharge parameters can account for geographic diversity at the national scale and provide a new data-based layer of information for sophisticated power system models to support energy transition.

\end{abstract}




\end{frontmatter}

\section{Introduction}

This study addresses the research question: \emph{What is the potential energy capacity and duration of heating flexibility from building fabric heat capacity?} A data-driven method for quantifying this heating flexibility duration at high spatial resolution is introduced and validated using Britain as a case study. 

Residential heating accounts for a significant share of greenhouse gas emissions for countries with cold climates: in the United Kingdom (UK), heating homes caused 17\% of emissions in 2019 \citep{HeatandBuildings2021}. Replacing gas boilers with electric heat pumps is expected to be the dominant strategy for heating decarbonization in the UK \citep{BuildingsUKClimateChangeCommittee2020}. Based on the 2018 and 2019 generation mix in Britain, making this switch reduces greenhouse gas emissions by at least 65\% \citep{Lizana2023}. 

Despite the carbon reductions that heat pumps offer, they pose a significant challenge to the power system. Electrifying residential heating is expected to significantly increase electricity demand, particularly at peak times. Heating all homes in Britain with heat pumps in 2019 would have added 42.3 GW of peak demand, increasing total peak electricity demand by 78\% \citep{Halloran2024}, and in a cold year in the 2020s, this figure could increase to 78 GW \citep{Watson2023}. 

For this reason, there has been increasing interest in the potential for heating flexibility, that is, shifting the consumption of heat pumps to reduce peak demand or achieve other benefits for the power system. \citet{Watson2023} show that if 100\% of hot water demand and 20\% of space heating demand were flexible, peak heat pump demand could be reduced by 16\%. Despite reported benefits, the large-scale quantification and management of heating flexibility capacity and duration, that is, how long homes in different regions turn their heating off before they become too cold, is a crucial challenge. 

Different approaches have been used to model the interaction between heating flexibility and the wider power system. They can be classified into three groups: detailed transient simulation models, reduced-order models, and steady-state models \citep{Li2021}. Each group has a different trade-off between building system model detail, energy system model detail, and the diversity of buildings represented.

Transient simulation models (also known as dynamic simulation models or white-box models)  are ``fundamentally based on conservation of energy, mass, and momentum" \citep{Li2021} and include detailed information about building structure, occupancy schedules, and heating systems. These models can represent a variety of building types, from a single archetype of a detached Irish house \citep{Bampoulas2021} to 144 configurations of Danish single-family homes \citep{Johra2019b}. However, transient simulation models tend to be too computationally intensive to couple with energy system models. A few studies \citep{LeDreau2016,Johra2019b} use dynamic electricity prices to account for building interactions with the power system. Similarly, \citet{Langevin2021} use net load (hourly load minus renewable generation) in 22 regions in the United States as a proxy for marginal electricity costs and quantify change in annual electricity use and daily net peak load from energy efficiency and demand response measures. More commonly, transient simulation models are used to quantify flexibility metrics of interest such as solar photovoltaic self-consumption \citep{Bampoulas2021,Reynders2013}; peak heating demand \citep{Reynders2013}; flexibility energy capacity \citep{Bampoulas2021,LeDreau2016,Reynders2017,Arteconi2019}, power capacity \citep{Reynders2017,Arteconi2019}, and duration \citep{LeDreau2016,Reynders2017,Arteconi2019}; flexibility recovery time \citep{Arteconi2019}; and thermal energy storage efficiency \citep{Bampoulas2021,Reynders2017}. 

To enable integration with power system models, reduced-order models (also called lumped-parameter, resistance-capacitance (RC), or gray-box models) represent only the main building components, decreasing computational complexity compared to transient simulation models  \citep{Li2021}. These reduced-order models represent the temperature of a small number of components, ranging from 2 to 3 \citep{HedegaardBalyk2013}, 4 \citep{Heinen2017}, and 5 thermal masses \citep{Patteeuw2015a,Patteeuw2015b}. These studies trade off between the details of the power system model and the diversity of building types considered. At one extreme, \citet{HedegaardBalyk2013} represent 10 archetypes of Danish single-family homes in a linear energy system investment optimization model considering five representative weeks at hourly temporal resolution. In contrast, \citet{Heinen2017} consider only 2 building archetypes (new and existing detached houses in Dublin, Ireland), and include hourly weather data for an entire year at a time in their electricity system planning model. Similarly, \citet{Patteeuw2015a} only include a single building archetype due to the computational complexity of the mixed-integer linear unit commitment and economic dispatch problem power system model. \citet{Patteeuw2015b} later consider 36 building types and 3 heating systems in this model, but each combination is considered individually due to computing constraints. 


The simplest heating flexibility models are thermal storage models based on a steady-state approach. These models limit the charging and discharging power of thermal storage as well as its state of charge, usually represented as an internal temperature within a range of comfortable temperatures. Self-discharge heat losses to the environment are commonly included based on a thermal time constant that includes a single thermal mass and a single thermal resistance \citep{Mathieu2015,Xue2014,Chen2014,Bloess2019,Zeyen2021}. As power system models increase in complexity \citep{Kotzur2021}, considering higher spatial and temporal resolution, thermal storage models offer an opportunity to include the variety of the building stock in these models without compromising computational tractability. Despite the opportunity to represent a large diversity of buildings, many of these models do not account for heterogeneity in the building stock in large geographies \citep{Chen2014,Bloess2019,Zeyen2021} while others only consider a few building parameter variations \citep{Xue2014,Hedegaard2012}. Considering 7 building archetypes to represent the German building stock, \citet{Papaefthymiou2012} use a transient simulation model to derive the parameters of a thermal storage model. \citet{Mathieu2015} consider regional weather differences to quantify flexibility potential from heat pumps and other thermostatic loads for 13 climate zones in California, but they assume a uniform distribution of thermal parameters for the entire state.

Despite the variety of approaches, none accurately represents the diversity of a national building stock at high spatial resolution and its flexibility duration in complex power system models. To overcome this challenge, a recent study explores the use of energy performance certificate (EPC) data. \citet{Canet2023} use a thermal storage model to reflect the spatial diversity in heating flexibility potential across geographies using an EPC-based method using national data at high spatial resolution in England and Wales. Because of its basis in EPCs, this method is likely to underestimate heating flexibility duration in the current residential building stock. Previous works have shown that EPCs are based on standard occupancy procedures that overestimate energy consumption \citep{Lizana2018b,Serrano-Jimenez2019} and assume default heating losses based on the year of construction that fail to capture retrofit improvements to the building fabric \citep{Ahern2020}. Moreover, comparing EPC-modeled energy use with smart meter-measured gas use in British households, \citet{Few2023} find that differences between the two for lower bands persist even in homes where occupant behavior matches EPC assumptions. These differences are likely to be particularly pronounced in rural areas, where \citet{Canet2022} found that an EPC-based method overestimated heating demand compared to gas consumption by a median of 61\% in the most rural LSOAs in England and Wales. Thus, although many methods have been applied to quantify heating flexibility, none accurately represent the geographic diversity of a national building stock with sufficiently low computational requirements to be incorporated with complex power system models (\textit{research gap 1}). Moreover, the only study that represents the variety of flexibility duration potential in British homes is likely underestimating that potential because of its basis in EPCs \citep{Canet2023}, which are known to overestimate heating demand (\textit{research gap 2}). 

This paper presents a novel heating consumption-based method for quantifying heating flexibility duration in the current building stock at high spatial resolution. The approach integrates high spatial resolution heating consumption data, temperature, and building size data to capture the geographic diversity of the national building stock. The method was tested in Britain and validated by comparing the results with the EPC-based method from \citet{Canet2023} and a temperature-based method using heat pump trial data from 742 homes \citep{EoHData2023}.

While this method can be applied to buildings in different sectors with different heating fuels, the scope of this case study only includes the residential sector. Because of the availability of high spatial resolution data on heating fuel consumption in Britain, this case study quantifies the flexibility duration potential of gas-heated homes. Therefore, the flexibility duration of homes in regions that are not on the gas grid is not quantified.

This study has two primary research contributions:
\begin{itemize}
    \item Introducing a method for quantifying national heating flexibility potential at high spatial resolution based on annual heating consumption data and historical temperature data (\textit{addressing research gap 1}). The software for this method is available under an MIT license on GitHub as \href{https://github.com/clairehalloran/GeoHeatFlex}{GeoHeatFlex}.
    \item Characterizing Britain's residential heating flexibility potential at high spatial resolution to effectively support power system planning (\textit{addressing research gap 2}). This dataset is openly available on the \href{https://ora.ox.ac.uk/objects/uuid:76f9bb1d-6d1b-4ffd-a238-8eb54f1df69e}{Oxford Research Archive} \citep{HalloranFlexPotential2024}.

\end{itemize}

This paper is structured as follows. First, the methods and case study data are described in Section \ref{sec:methods and data}. Second, the results are divided into Section \ref{sec:heat capacity and heat losses} describing the spatial variation in heat capacity and heat loss rate, the two most important variables in the model; Section \ref{sec:validation results} discussing the model validation; and Section \ref{sec:case study results} presenting the results for the case study of gas-heated British homes. Finally, the conclusions are drawn in Section \ref{sec:conclusions}, highlighting the practical implications of this approach.

\section{Data and methods}\label{sec:methods and data}
\begin{figure*}[!htp]
\centering
\includegraphics[width=2\columnwidth]{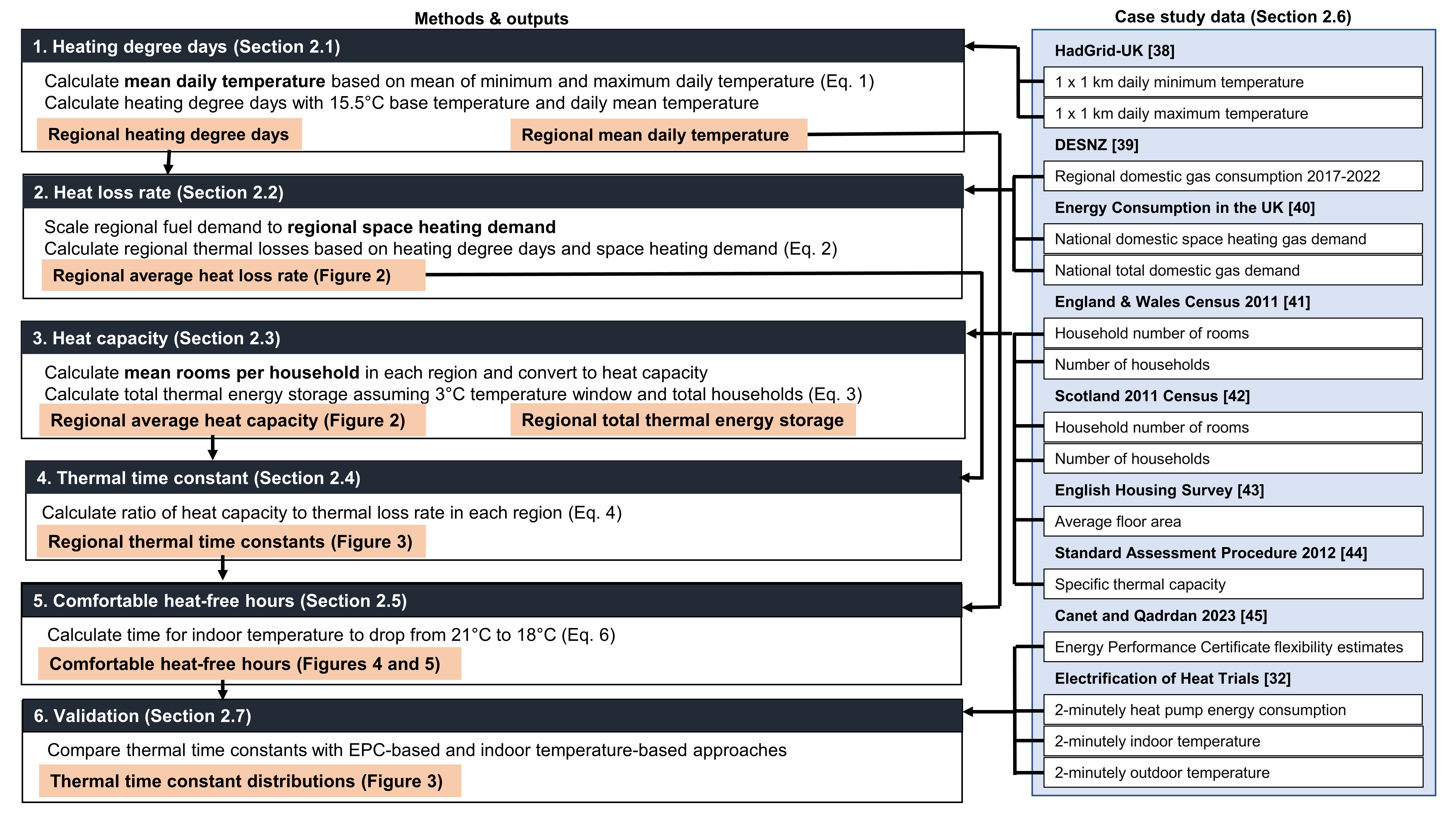}

\caption{Summary of methods and case study data used in this paper. Outputs for each step are shown in orange boxes, and intermediate outputs for each step are shown in bold.}
\label{fig:methods}
\end{figure*}

Figure \ref{fig:methods} summarizes the methods and data used in this paper. Sections \ref{sec:HDDs} through \ref{sec:methods heat-free hours} introduce the method. Section \ref{sec:case study} introduces the British case study and demonstration input data, and Section \ref{sec:validation} discusses the validation of this heating consumption-based method.

\subsection{Heating degree days}\label{sec:HDDs}
Heating degree days (HDDs) are calculated based on gridded historical minimum and maximum temperature observations. Following \citet{Kendon2021}, the mean daily temperature for each point is calculated based on the maximum and minimum temperature observations:

\begin{equation}
    T^{mean}=\frac{T^{max}+T^{min}}{2}
\end{equation}

HDDs are calculated based on the resulting daily mean temperature $T^{mean}$ assuming a base temperature of 15.5ºC. Gridded annual HDDs are assigned to each region based on the location of the representative point of each region. If the representative point is not within the gridded range, the mean number of HDDs among all neighboring regions is assigned.

\subsection{Heat loss rate}

The average heat loss rate for residential buildings in each region $r$ is calculated using a top-down methodology from \citet{Chen2014} based on annual heating consumption $Q_r$ and HDDs $HDD_r$:
\begin{equation}
    H_r=\frac{Q_r}{24\sum HDD_r}
\end{equation}
HDDs are multiplied by 24 hours to give heat loss rate $H_r$ in kW/ºC. 

\subsection{Heat capacity}

The average heat capacity $C_r$ in each region is calculated based on the average number of rooms per household. This value is converted to an estimated floor area and finally to heat capacity using a heat capacity per floor area value.

The average heat capacity and the number of households in each region $A_r$ is used to calculate the total thermal energy storage in the building fabric for each region. For a flexibility window $\Delta T^{flex}$, the total thermal energy storage capacity $\bar{e_r}$ in each region is:

\begin{equation}
    \bar{e_r} = C_r A_r \Delta T^{flex}
\end{equation}

\subsection{Thermal time constant}

Spatial variation in the duration of heating flexibility is accounted for using the average time constant $\tau_r$ for each region $r$. This lumped thermal parameter reflects how quickly heat is lost from residential buildings in each region, depending on the temperature difference between indoor and outdoor air. The time constant is the ratio of the total heat capacity $C_r$ to the total heat losses $H_r$ for each region: 
	\begin{equation}\label{eq:time constant}
		\tau_r=\frac{C_r}{H_r}
	\end{equation}

\subsection{Comfortable heat-free hours}\label{sec:methods heat-free hours}
 
Newton's law of cooling relates the time constant $\tau_r$ with the indoor air temperature at a given time $t$ to the outdoor air temperature $T^{out}_r$ (assumed constant in time) in each region $r$ and initial indoor air temperature $T^{in}_{r,0}$:
\begin{equation}\label{eq:newton cooling}
    T^{in}_{r,t}=T^{out}_r+(T^{in}_{r,0}-T^{out}_r)e^{-t/\tau_r}
\end{equation}

Assuming an indoor temperature range from 18 to 21ºC, the number of comfortable heat-free hours $t^{c}_r$ in each region $r$ is calculated as:

\begin{equation}\label{eq:comfortable heat-free hours}
    t^{c}_r=-\tau_r\ln{\frac{18-T^{out}_r}{21-T^{out}_r}}
\end{equation}

The number of heat-free hours in each region is calculated based on a variety of outdoor temperatures. To provide an even comparison of the residential building stock across Britain, comfortable heat-free hours are calculated using a uniform outdoor temperature $T^{out}_r$ of 5ºC in every region $r$. To account for regional variation in typical winter temperatures, the number of heat-free hours is also calculated for different percentiles of historical regional daily winter temperatures from 2010 to 2022. This time period was selected to capture both changing winter patterns under climate change as well as extreme cold periods in winter 2010/11.

\subsection{Case study data}\label{sec:case study}

The heating consumption-based method for calculating heating flexibility duration introduced in Sections \ref{sec:HDDs} through \ref{sec:methods heat-free hours} is applied to a case study of gas-heated homes in Britain. For each step, values are calculated for each 2011 Lower Level Super Output Area (LSOA) in England and Wales and each 2011 Data Zone (DZ) in Scotland. LSOAs are regions with an average of 1,500 residents and a minimum of 1,000 residents, while DZs are slightly smaller, with populations between 500 and 1,000 \cite{DESNEZSubnationalMethods}. There are 34,753 LSOAs in England and Wales and 6,505 DZs in Scotland. The boundaries for these regions are available from \citet{LSOA2011} and \citet{DZ2011}.

Historical minimum and maximum temperature observations at 1 x 1 km resolution from HadUK-Grid \citep{HadUK-Grid} are used to calculate HDDs and regional daily winter temperature percentiles. Annual HDDs are calculated based on the gas heating season the annual metered domestic gas demand data, which is typically from May of each year to the following May \citep{DESNEZSubnationalMethods}. 

Heating demand for each region is based on annual metered domestic gas consumption from the Department for Energy Security and Net Zero (DESNZ) \citep{LSOAGas2021}. Because the gas data methodology changed in 2017, only data from 2017 to 2021 is used  \citep{DESNEZSubnationalMethods}. Total gas consumption is scaled to space heating demand in each region based on the national share of domestic gas consumption used for space heating for each year \cite{ECUK2022-EndUse}.  Heat losses are based on total space heating gas consumption and HDDs from 2017 to 2021.

The average number of rooms in homes in each region is obtained from the 2011 censuses \citep{RoomsEW2011,RoomsS2011}. Based on the mean number of rooms in English homes and the mean floor area of 94 $m^2$ from the English Housing Survey \citep{EHSSize2019}, the average number of rooms is converted to floor space assuming 17.6 $m^2$ of floor area per room. The Standard Assessment Procedure medium heat capacity value of 250 kJ/m²ºC is used to convert floor area to heat capacity \citep{SAP2012}.

\subsection{Validation}\label{sec:validation}
To validate this heating consumption-based method, the regional thermal time constants are compared with those obtained from 2 different data sources: the EPC-based method from \citet{Canet2023} and an indoor temperature-based method using 2-minutely temperature and energy use data from a sample of 742 homes in the Electrification of Heat trial \citep{EoHData2023}.

\subsubsection{EPC-based method}

Thermal building characteristics calculated based on the EPC-based method \citep{Canet2023} are available for each dwelling category (flats, terraced, semi-detached, and detached) and heating system in each LSOA in England and Wales \citep{CanetData2023}. Considering only gas-heated households, the time constant for gas-heated homes in each dwelling category in each LSOA is calculated based on the thermal capacity and thermal losses using Equation \ref{eq:time constant}. Since \citet{Canet2023} present thermal losses for both the current building stock and potential retrofit, current and retrofit time constants are calculated.

\subsubsection{Indoor temperature-based method}

The Electrification of Heat trial includes 742 households in Southeast England, Northeast England, and Scotland with heat pumps. A cleaned interim dataset including 2-minutely internal temperature, external temperature, and heat pump energy output from November 2020 to August 2022 for 740 of these homes is available \citep{EoHData2023}. 

The time constants of the homes included in this trial were calculated using an indoor temperature-based method. This approach uses an exponential fit of the internal temperature decrease when the heat pump, as well as any boiler or backup heater, was off. The internal temperature drop was fit using the following equation, based on Equation \ref{eq:newton cooling}:

\begin{equation}
    T^{in}(t)=T^{out}_{mean}+(T^{in}-T^{out}_{mean})e^{-t/\tau}
\end{equation}

Because Equation \ref{eq:newton cooling} assumes constant outdoor temperature, the mean  outdoor temperature is used to fit this equation. 

Only data from time periods during the heating season (from October to April) and with both internal and external temperature data availability are considered to avoid data from warm days. Moreover, three data pre-processing steps were applied. First, periods were selected when the heating was off for at least 90 minutes, and second, time periods with sudden, large drops in temperature (more than 5ºC in 2 minutes) were excluded based on the assumption that an external door or window was opened. Third, because Equation \ref{eq:newton cooling} assumes constant outdoor temperature, only time periods with less than 2ºC variation in outdoor air temperature were considered. All homes have at least 1 time period that fits these criteria. For homes with multiple time periods that fit these criteria, the mean of the fitted $\tau$ values for each period is taken as the time constant for that house.




\section{Results and discussion}\label{sec:results}
This section discusses the results and their implications for heating flexibility in Britain. First, the regional variation in heat capacity and heat loss rates is presented in Section \ref{sec:heat capacity and heat losses}. Then, Section \ref{sec:validation results} presents the time constants calculated using the heating consumption-based approach and validates them with two other methods: the EPC-based method from \citet{Canet2023}  and a temperature-based method using heat pump trial data from 742 homes \citep{EoHData2023}. Finally, Section \ref{sec:case study results} explores the implications of these findings for flexibility duration in Britain using uniform and spatially varied outdoor temperatures.

\subsection{Heat capacity and heat loss rate}\label{sec:heat capacity and heat losses}

\begin{figure}
\centering
\includegraphics[width=\columnwidth]{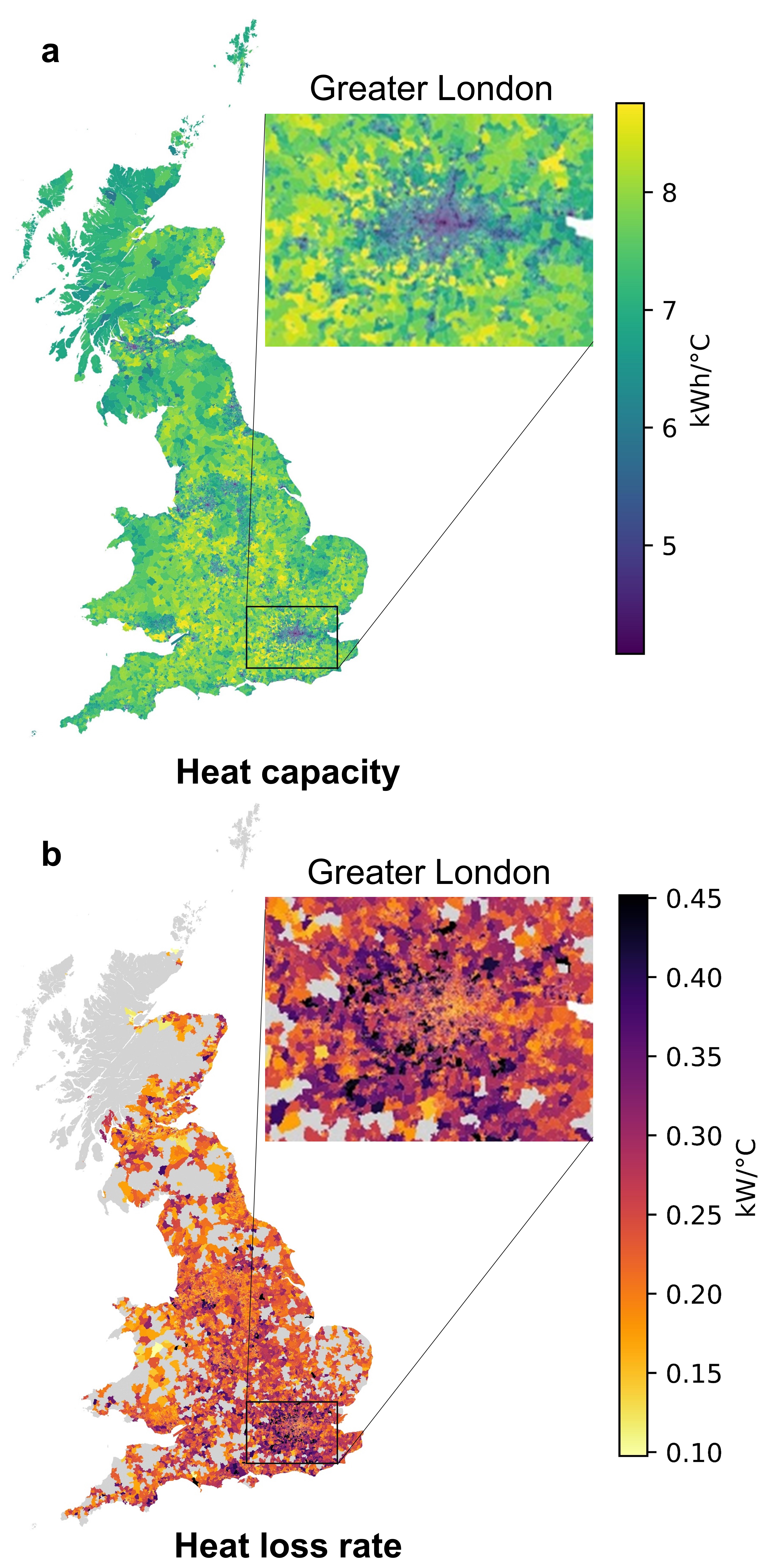}

\caption{Maps of (a) mean heat capacity in each region and (b) mean heat loss rate in houses heated with gas in each region. Inset maps display details for the Greater London area. Domestic gas consumption data is unavailable for regions shown in gray. Lower layer Super Output Area (LSOA) boundaries \citep{LSOA2011} and Data Zone (DZ) boundaries \citep{DZ2011} licensed under the \href{https://www.nationalarchives.gov.uk/doc/open-government-licence/version/3/}{Open Government Licence v.3.0}. LSOA boundaries contains OS data © Crown copyright and database right 2023. DZ boundaries copyright Scottish Government, contains Ordnance Survey data © Crown copyright and database right 2021.}
\label{fig:building characteristics}
\end{figure}

Regional time constants are the ratio of heat capacity to heat loss rate. Buildings with higher heat capacity (kWh/ºC) can store more thermal energy in the building fabric and maintain the building temperature for longer while the heating is off. In this study, heat capacity is proportional to building size based on the mean number of rooms per household. Buildings with higher heat loss rates (kW/ºC) are worse-insulated and lose more energy to their surroundings when the outdoor temperature is below the indoor temperature. In this study, higher heat loss rates reflect higher gas consumption per HDD. The regional values of heat capacity and heat loss rate are illustrated in Figure \ref{fig:building characteristics}, which reveals significant geographic variation.

Figure \ref{fig:building characteristics}a displays the mean heat capacity in each region. The median heat capacity in each region is 6.5 kWh/ºC; however, heat capacity varies greatly among regions, ranging from 10 kWh/ºC in rural regions and heat capacity to 3 kWh/ºC in densely populated urban regions. These figures reflect the diversity in the size of dwellings in Britain. Assuming a 3ºC temperature flexibility window, the total thermal energy storage capacity in the British housing stock, including both gas- non-gas-heated homes, is 500 GWh$_{th}$. This figure is not directly comparable to electricity storage because the amount of delayed electricity consumption depends on the temperature-dependent COP of the heat pumps. For an illustrative cold temperature COP value of 2.5~\cite{EOHInterim}, this thermal energy storage capacity is equivalent 200 GWh of electricity storage. Note that this figure is an order-of-magnitude estimate calculated assuming a uniform room size and specific heat capacity across the housing stock.

Figure \ref{fig:building characteristics}b displays the mean heat loss rate in gas-heated dwellings in each region. Regions with too few houses on the gas grid to be included in the metered demand data are in gray. Median heat losses per region are 0.21 kW/ºC, but there is significant regional variation from 0.12 kW/ºC at the 1st percentile to 0.41 kW/ºC at the 99th percentile. High heat losses in the outer regions of Greater London stand out. Unlike heat capacity, there is not a clear spatial pattern in heat losses that can be predicted from population density. This underlines the importance of using the method presented in this paper to analyze regional heat losses.

\subsection{Model validation}\label{sec:validation results}

\begin{figure}[!htp]
\centering
\includegraphics[width=\columnwidth]{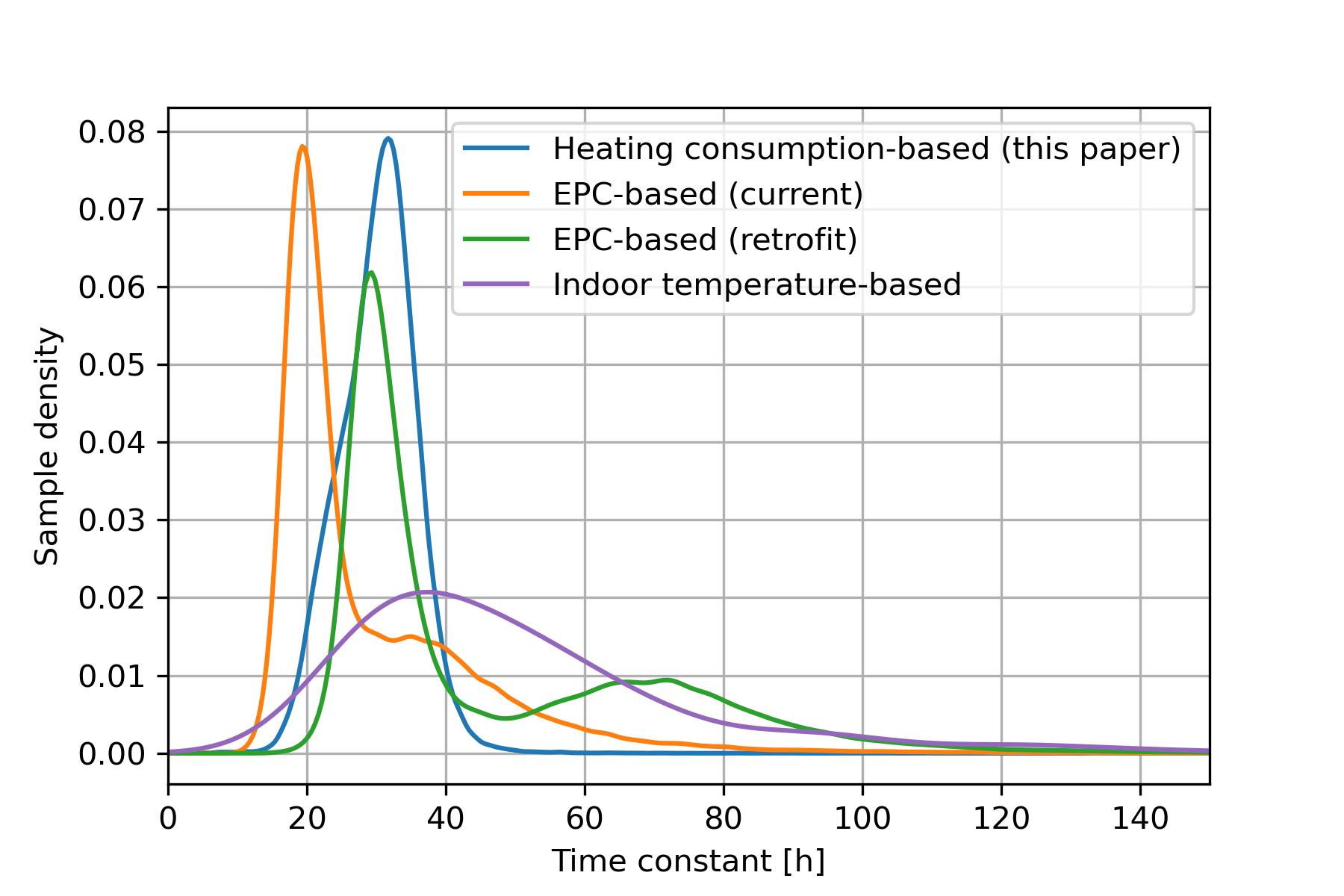}

\caption{Comparison of thermal time constants for the British housing stock based on the heating consumption-based approach introduced in this paper, the EPC-based method for current and retrofit housing stock, and an indoor temperature-based method using data from heat pump trials in 742 homes.}
\label{fig:KDE time constant histogram}
\end{figure}

The heating consumption-based method introduced in this paper calculates the thermal time constant (h) for each region on the ratio of thermal capacity to heat loss rate presented in Section \ref{sec:heat capacity and heat losses}. In this section, the heating consumption-based thermal time constants are compared with time constants obtained from the EPC-based approach and the indoor temperature-based method. Figure \ref{fig:KDE time constant histogram} illustrates the distribution of time constant value for each approach. The results demonstrate that time constants calculated based on the heating consumption-based method are consistent with those calculated for the British housing stock using other methods as follows.

The blue line shows the results from the heating consumption-based method using national data at high spatial resolution presented in this paper. The median time constant value is 30.6 hours. 

The orange and green lines illustrate the distribution of time constants from the EPC-based method for the current building stock in England and Wales as well as the retrofit potential of the building stock. The similar magnitude of time constants obtained from the heating consumption-based and EPC-based methods verifies the soundness of the heating consumption-based results. However, because EPCs overestimate heating losses \citep{Ahern2020,Few2023}, the EPC-based method for the current building stock results in a lower median time constant of 22.8 hours than the heating consumption-based method. In contrast, the distribution of time constants from the EPC-based method for retrofit potential, with a median value of 33.8 hours, closely matches the distribution of heating consumption-based time constants. This agreement may reflect building retrofits that have taken place since the last EPC assessment, which is updated every 10 years, as well as energy-saving occupant behaviors that EPCs do not consider. Because the EPC-based method represents each dwelling category in each region separately, there are smaller peaks in the EPC-based time constant distributions at higher time constant values. These peaks represent high heat capacity detached dwellings with low heat loss rates.

Finally, the purple line shows the distribution of time constants obtained with the indoor temperature-based method using a sample of 742 homes in the Electrification of Heat trial. These results further support the validity of the consumption-based time constant values. The median time constant for these homes is 44.4 hours, which is almost double the median heating consumption-based regional time constant value of 30.6 hours. This difference can be explained by the fact that compared to the national average, the Electrification of Heat trial includes a higher proportion of detached and semi-detached houses, which tend to have higher floor areas and higher thermal capacities, and a lower proportion of flats, which have lower floor areas and lower thermal capacities \citep{EoHInstall2022}. In addition, properties with a heat demand that was too high were excluded from the trial, so this data is representative of a well-insulated building stock with low heat losses. 

\subsection{Case study: British residential flexibility duration}\label{sec:case study results}

\begin{figure*}[!htp]
\centering
\includegraphics[width=2\columnwidth]{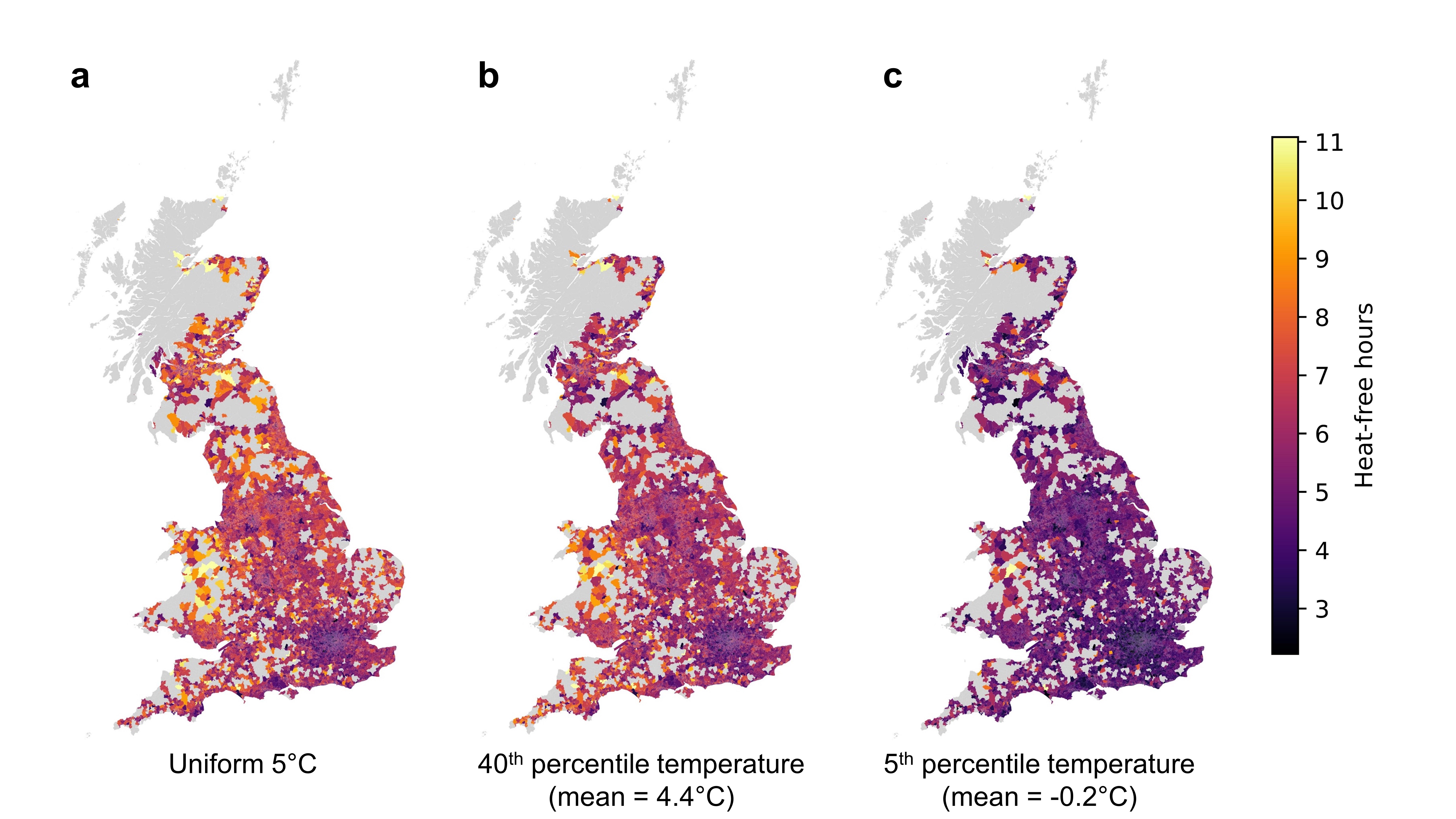}

\caption{Maps of average comfortable heat-free hours in each region, for an indoor window of 18ºC to 21ºC with (a) a uniform outdoor temperature of 5ºC and (b) 40th percentile and (c) 5th percentile regional daily winter temperature for 2010 to 2022. Map (a) displays the regional variation in flexibility duration due to building stock differences, while maps (b) and (c) demonstrate the effect of regional temperature differences on heat-free hours. Domestic gas consumption data is unavailable for regions shown in gray. Lower layer Super Output Area (LSOA) boundaries \citep{LSOA2011} and Data Zone (DZ) boundaries \citep{DZ2011} licensed under the \href{https://www.nationalarchives.gov.uk/doc/open-government-licence/version/3/}{Open Government Licence v.3.0}. LSOA boundaries contains OS data © Crown copyright and database right 2023. DZ boundaries copyright Scottish Government, contains Ordnance Survey data © Crown copyright and database right 2021.}
\label{fig:heat-free hours map}
\end{figure*}

Due to the geographical diversity in the current residential building stock and outdoor temperature, flexibility duration varies considerably among regions in Britain. Figure \ref{fig:heat-free hours map} displays the number of comfortable heat-free hours in each region for an indoor temperature window from 18ºC to 21ºC and various outdoor temperatures.

Figure \ref{fig:heat-free hours map}a displays the number of comfortable heat-free hours in each region for an indoor temperature window from 18ºC to 21ºC with a uniform outdoor temperature of 5ºC. The median number of heat-free hours is 6.4, with an interquartile range of 5.5 to 7.0 hours. A few areas in central Wales, northern England, and Scotland stand out with higher-than-average heat-free hours. Despite regional differences in heat capacity and heat losses in Greater London (as shown in the inset maps in Figure \ref{fig:building characteristics}), the region has uniformly low heat-free hours. Both small, well-insulated flats in central London, with low heat capacity and low heat losses and large, poorly-insulated detached houses with a high heat capacity and high heat losses in the surrounding areas result in similarly low heating flexibility duration. Notably, this work does not consider heating flexibility power, which determines the total amount of energy shifted along with flexibility duration. Regions with small, well-insulated flats are likely to have lower total heat pump capacity than regions with large, poorly insulated detached houses. Therefore, central London is expected to have lower flexibility power capacity than the surrounding regions despite having similar flexibility duration.

Figure \ref{fig:heat-free hours map}b displays heat-free hours for the 40th percentile daily winter temperature in each region. The regional mean value of the 40th percentile daily winter temperature is 4.4ºC. Comparing Figure \ref{fig:heat-free hours map}b with Figure \ref{fig:heat-free hours map}a, which has a uniform outdoor temperature close to the mean temperature in (b), demonstrates the effect of accounting for regional differences in outdoor temperature on heat-free hours. In many central and northern regions, heat-free hours decrease, and they increase heat-free hours in southern coastal regions. Figure \ref{fig:heat-free hours map}c demonstrates that for 5th percentile temperatures, flexible heat-free hours reduce to less than 5 hours in the vast majority of regions.

\begin{figure}[!htp]
\centering
\includegraphics[width=\columnwidth]{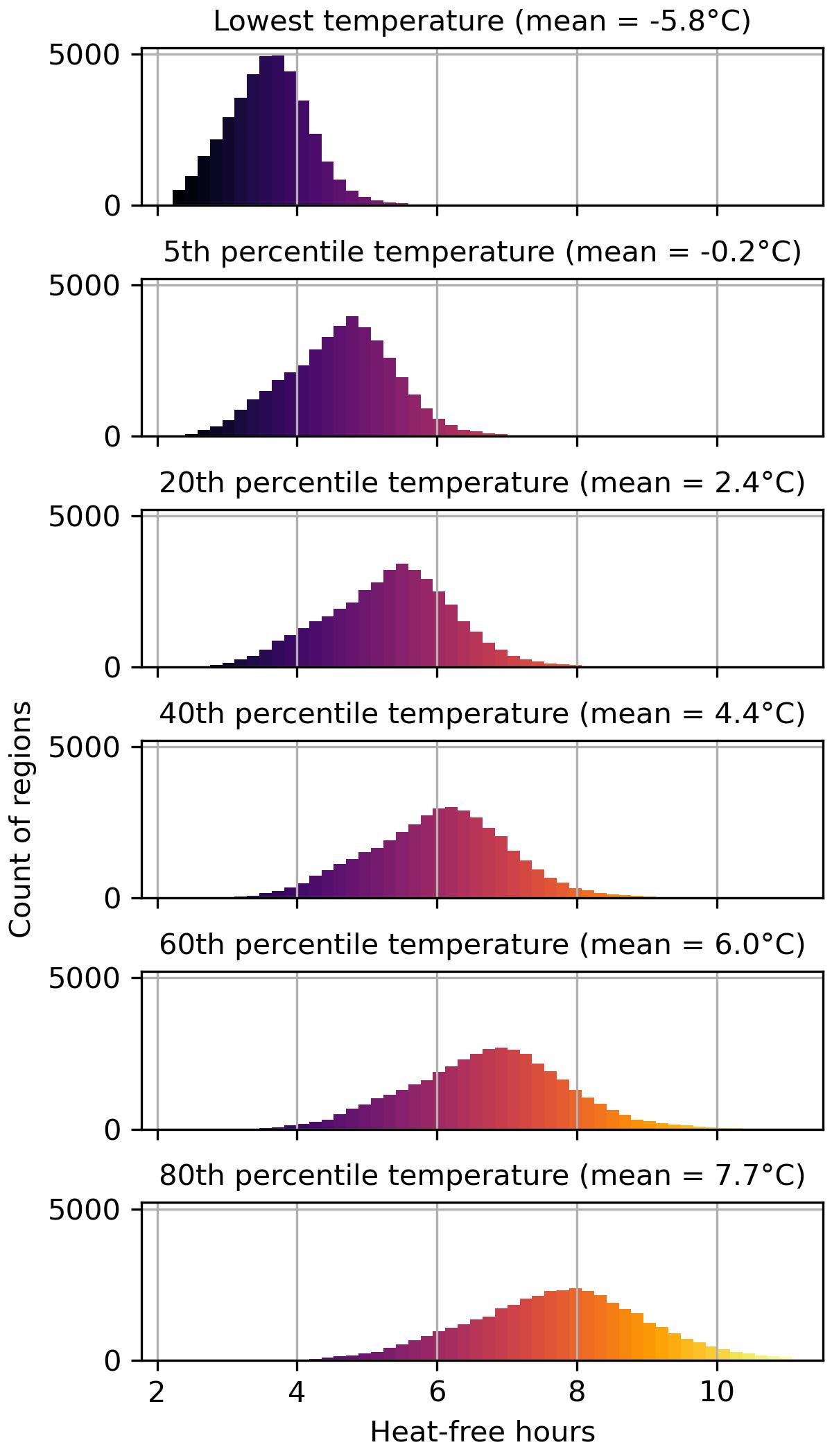}

\caption{Histogram of average comfortable heat-free hours in each region for various outdoor temperature percentiles during winter 2010 to 2022. Mean temperatures in each subfigure title are the mean regional temperature at each percentile. Note that the colormap matches Figure \ref{fig:heat-free hours map}.}
\label{fig:time constant histogram}
\end{figure}

Accounting for daily variation in winter temperatures demonstrates the impact of outdoor temperature on heating flexibility duration. Figure \ref{fig:time constant histogram} shows the distribution of comfortable heat-free hours for each region at different historical temperature percentiles during winters from 2010 to 2022.  

At lower outdoor temperatures, the median number of heat-free hours in each region drops: for the 80th percentile temperature, the median is 7.8 hours and for the 5th percentile temperature, the median is 4.7 hours. The range of regional heat-free hours also decreases as the outdoor temperature drops. For the 80th percentile temperature, the interquartile range is 6.9 to 8.6 hours, and for the 5th percentile temperature, the interquartile range is 4.1 to 5.1 hours. This range decrease means that regions with higher-than-median thermal time constants will experience larger flexibility duration drops in cold weather than regions with lower-than-median thermal time constants. This increased sensitivity to outdoor temperature will limit the ability of heating flexibility from regions with high thermal time constants to support the power system during long cold snaps.

Although the current building stock offers significant flexibility duration during typical winter days, flexibility duration drops significantly during extreme cold events. For the lowest daily winter temperature from 2010 to 2022, with a mean value of -5.8ºC, the median flexibility duration was just 3.6 hours with an interquartile range of 3.2 to 4.0 hours. This decreased duration will limit the ability of heating flexibility in the current building stock to ensure power system reliability during extreme cold events, which are predicted to be a major challenge for the electricity system in the future with electrified heating. This finding suggests that longer-duration forms of flexibility will be necessary to maintain power system reliability during extreme cold events.

\section{Conclusions}\label{sec:conclusions}

This paper aims to quantify the energy capacity and duration of heating flexibility available from the existing building stock at high spatial resolution to support power system planning. While many existing methods model the interaction of heating flexibility with the power system, there is a need to represent the geospatial diversity of the building stock in a heating flexibility model that can be incorporated into complex power systems models. To this end, a method for assessing heating flexibility potential based on thermal time constants calculated from historical heating energy consumption, temperature, and building size is introduced. This method is applied to a case study of gas-heated homes in Britain and validated by comparing thermal time constants with those obtained from energy performance certificates (EPCs) and a heat pump trial. The number of comfortable heat-free hours without heating was presented for a constant outdoor temperature of 5ºC as well as different percentiles of regional winter temperatures. Based on the results, the following conclusions can be drawn:

The proposed method produces high spatial resolution thermal time constant values that are broadly consistent with those found using EPC-based and indoor temperature-based methods. The heating consumption-based thermal time constant values are higher than those obtained using current EPC values (median value of 30.6 hours vs. 22.8 hours), which is expected because EPCs tend to overestimate heating demand. Compared to heat pump trial data with a median of 44.4 hours, the values obtained from this paper's method are slightly lower. This difference is explained by the disproportionate representation of detached and semi-detached houses in the trial, as well as the exclusion of homes with high heating demand. Thus, the proposed consumption-based method produces more accurate thermal time constant estimates for the current housing than the EPC-based method and at higher spatial resolution than the indoor temperature-based method.

In the case study of Britain, significant heating flexibility energy capacity and duration are identified in the existing housing stock. The total thermal energy storage capacity in residential buildings of both gas and non-gas-heated homes is 500 GWh$_{th}$ for a 3ºC temperature flexibility window. This is equivalent to 200 GWh of electricity storage for a cold weather COP value of 2.5. The housing stock also has significant heating flexibility duration potential: assuming a uniform 5ºC outdoor temperature and indoor temperature range from 18 to 21ºC, regions had a median of 6.4 comfortable heat-free hours and an interquartile range of 5.5 to 7.0 heat-free hours. However, heating flexibility duration varies significantly among regions. Due to low thermal capacity in central London and high heating losses in the surrounding home counties, there is relatively low flexibility duration in Greater London and the Southeast. High flexibility duration is identified in some rural areas due to higher heat capacity and lower heat losses. 

Accounting for the impact of historical outdoor temperatures demonstrates that extreme cold events can drastically decrease flexibility duration compared to typical winter temperatures. For the 20th percentile of daily winter temperatures, the median regional heat-free hours is 5.9, with an interquartile range from 5.2 to 6.5 hours. At these temperatures, heating flexibility can compete with the typical 4-hour duration of lithium-ion batteries during the heating season. However, for the lowest daily winter temperatures in each region from 2010 to 2022, this heating flexibility duration nearly halved to a median of 3.6 hours with an interquartile range of 3.2 to 4.0 hours. This temperature-driven decrease in flexibility duration will limit the ability of heating flexibility to contribute to power system reliability during extreme cold events.

The results demonstrate how this novel consumption-based method can characterize the heating flexibility potential of the current building stock at high spatial resolution. This approach provides building stock data that can be incorporated into sophisticated power system planning models. Future work will evaluate the role of heating flexibility in generation, storage, and transmission expansion planning for bulk power systems with high shares of renewable generation. It should be noted that this study quantifies heating flexibility duration and does not consider flexibility power, which depends on the heat pump sizing and building efficiency. Because of the uncertainty about the spatial distribution of these characteristics, this topic will be addressed in future studies.

\section*{CRediT authorship contribution statement}
\noindent \textbf{Claire Halloran:} Conceptualization, Methodology, Software, Validation, Writing - Original draft, Visualization. \textbf{Jesus Lizana:} Writing - Review \& Editing. \textbf{Malcolm McCulloch:} Supervision.

\section*{Declaration of competing interest}
\noindent The authors declare that they have no known competing financial interests or personal relationships that could have appeared to influence the work reported in this paper.

\section*{Acknowledgments}
\noindent C.~Halloran acknowledges the support of the Rhodes Trust.



\bibliographystyle{elsarticle-num-names} 
\bibliography{references}





\end{document}